\documentclass[times,twocolumn]{aastex63}
\usepackage{amsmath}
\usepackage{amssymb}

\newcommand{\dpar}[2]{\frac{\partial#1}{\partial#2}}

\newcommand{\bracketfunc}[3]{\left(\frac{#1}{#2}\right)^{#3}}
\newcommand{\bracketfuncb}[3]{\left[\frac{#1}{#2}\right]^{#3}}

\newcommand{\mpl}{M_{\rm p}}
\newcommand{\mstar}{M_{\ast}}
\newcommand{\rp}{R_{\rm p}}
\newcommand{\rgap}{R_{\rm gap}}
\newcommand{\tgap}{t_{\rm gap}}
\newcommand{\tmig}{t_{\rm mig}}
\newcommand{\rhosurf}{\Sigma}
\newcommand{\vvec}{\vec{v}}
\newcommand{\vrad}{\rm{v}_{R}}
\newcommand{\vphi}{\rm{v}_{\phi}}
\newcommand{\grav}{\rm{G}}
\newcommand{\h}{h}
\newcommand{\hp}{h_{\rm p}}
\newcommand{\sigmagap}{\rhosurf_{\rm gap}}

\newcommand{\sigmaun}{\rhosurf_{\rm un}}
\newcommand{\sigmaunp}{\rhosurf_{\rm un,p}}
\newcommand{\rhill}{R_{\rm H}}

\newcommand{\omegak}{\Omega_{\rm K}}
\newcommand{\omegakp}{\Omega_{\rm K,p}}
\newcommand{\depsecond}{\delta_{\rm gap}^{\rm 2nd}}
\newcommand{\rgapsecond}{R_{\rm gap}^{\rm 2nd}}

\newcommand{\hco}{$^{13}$CO }
\newcommand{\cho}{C$^{18}$O }

\newcommand{\RED}[1]{\color{black} #1 \color{black}}
\newcommand{\REDD}[1]{\color{black} #1 \color{black}}
\received{\today}
\revised{\today}
\accepted{\today}
\submitjournal{ApJ}

\shorttitle{Model of a gap formed by a planet with fast inward migration}
\shortauthors{K.D. Kanagawa et al.}
\begin{document}

%\title{A fast inward migration of a planet within a protoplanetary disk: Observational signatures from gas morphology}
\title{Model of a gap formed by a planet with fast inward migration}

%\author[0000-0001-7235-2417]{Kazuhiro D. Kanagawa}
\author{Kazuhiro D. Kanagawa}
\affiliation{Research Center for the Early Universe, Graduate School of Science, University of Tokyo, Hongo, Bunkyo-ku, Tokyo 113-0033, Japan}

\author{Hideko Nomura}
\affiliation{National Astronomical Observatory of Japan, 2-21-1 Osawa, Mitaka, Tokyo 181-8588, Japan}

\author{Takashi Tsukagoshi}
\affiliation{National Astronomical Observatory of Japan, 2-21-1 Osawa, Mitaka, Tokyo 181-8588, Japan}

\author{Takayuki Muto}
\affiliation{Division of Liberal Arts, Kogakuin University, 1-24-2 Nishi-Shinjuku, Shinjuku-ku, Tokyo 163-8677, Japan}

\author{Ryohei Kawabe}
\affiliation{National Astronomical Observatory of Japan, 2-21-1 Osawa, Mitaka, Tokyo 181-8588, Japan}
\affiliation{The Graduate University for Advanced Studies (SOKENDAI), 2-21-1 Osawa, Mitaka, Tokyo
181-0015, Japan}
\affiliation{Department of Astronomy, The University of Tokyo, Hongo, Tokyo 113-0033, Japan}

\correspondingauthor{Kazuhiro D. Kanagawa}
\email{kazuhiro.kanagawa@utap.phys.s.u-tokyo.ac.jp}



\begin{abstract}
A planet is formed within a protoplanetary disk.
Recent observations have revealed substructures such as gaps and rings, which may indicate forming planets within the disk.
Due to disk--planet interaction, the planet migrates within the disk, which can affect a shape of the planet-induced gap.
In this paper, we investigate effects of fast inward migration of the planet on the gap shape, by carrying out hydrodynamic simulations.
We found that when the migration timescale is shorter than the timescale of the gap-opening, the orbital radius is shifted inward as compared to the radial location of the gap.
We also found a scaling relation between the radial shift of the locations of the planet and the gap as a function of the ratio of the timescale of the migration and gap-opening.
Our scaling relation also enables us to constrain the gas surface density and the viscosity when the gap and the planet are observed.
Moreover, we also found the scaling relation between the location of the secondary gap and the aspect ratio.
By combining the radial shift and the secondary gap, we may constrain the physical condition of the planet formation and how the planet evolves in the protoplanetary disk, from the observational morphology.
\end{abstract}


\section{Introduction} \label{sec:intro}
In a protoplanetary disk, a planet is formed and its orbital radius of the planet varies by gravitational interaction to the surrounding gas \citep[e.g.,][]{Lin_Papaloizou1979,Goldreich_Tremaine1980}, and its mass increases by the gas accretion onto the planet \citep[e.g.,][]{Bryden1999,Kley1999,D'Angelo_Kley_Henning2003,Tanigawa_Watanabe2002,Machida_Kokubo_Inutsuka_Matsumoto2010}.
Moreover, when the mass of the planet is massive enough, the planet opens a density gap along with its orbit and it migrates with the gap \citep[e.g.,][]{Lin_Papaloizou1986b,Edgar2007,Crida_Morbidelli2007,Durmann_Kley2015,Durmann_Kley2017,Kanagawa_Tanaka_Szuszkiewicz2018,Kanagawa2019}.
Outside of the gap, moreover, relatively large dust grains can be piled-up and a ring structure can be formed \citep[e.g.,][]{paardekooper2004,Muto_Inutsuka2009b,Zhu2012,Dong_Zhu_Whitney2015,Weber2018,Kanagawa_Muto_Okuzumi_Taki_Shibaike2018}.
Such gap/ring structures in protoplanetary disks can be considered as a signal of the planet formation.

Recent observations have revealed a large diversity of exoplanets including a close-in giant planet (Hot Jupiter) and Super-Earths \citep[e.g.,][]{Winn_Fabrycky2015}, and giant planets orbiting at large radii \citep[e.g.,][]{Hashimoto2012}.
Although the origin of the diversity is still not understood, it could be related to how the planets form and evolve within protoplanetary disks.
Thanks to e.g., the Atacama Large Millimeter/submillimeter Array (ALMA) and Subaru telescope, substructures such as rings, gaps, and spirals have been observed at protoplanetary disks \citep[e.g.,][]{Fukagawa2013,ALMA_HLTau2015,Akiyama2015,Momose2015,van_der_Plas_etal2017,Fedele_etal2017,Dong2018_MWC758,Long2018,Marel2019}.
From the depth and width of the observed gap, one can estimate the mass of the unseen planet embedded in the disk \citep[e.g.,][]{Kanagawa2015b,Kanagawa2016a,Rosotti_Juhasz_Booth_Clarke2016,Zhang_DSHARP2018}, if the gap is formed by the planet.
Moreover, recent observations have discovered point sources in the protoplanetary disks, PDS~70 \citep{PDS70_Keppler2018,PDS70_Muller2018}, TW~Hya \citep{Tsukagoshi2019}, which are candidates of the forming planet.
These observations enable us to know the presence of the planet in the present stage.
To reveal when and where the planet is formed, however, we need to consider how the planet evolves within the protoplanetary disk.
It is still an open question how the planet evolves within the protoplanetary disks, though it has been actively studied from a theoretical point of view \citep[e.g.,][]{Mordasini_Alibert_Benz_Klahr_henning2012,Ida_Lin_Nagasawa2013,Bitsch_Lambrechts_Johansen2015,Ida_Tanaka_Johansen_Kanagawa2018,Johansen_Ida_Brasser2019,Tanaka_Murase_Tanigawa2019}.

\cite{Meru_Rosotti_Booth_Nazari_Clarke2018} and \cite{Nazari2019} have investigated observational signatures of the planetary migration by focusing on locations of gas pressure bumps and dust rings.
\cite{Weber2019} have also investigated the effects of the migration on the location of the dust rings in the case of low viscosity.
%In the near future, however, the shape of the gap and the location of the planet could be directly observed by the gas emission.
\cite{Meru_Rosotti_Booth_Nazari_Clarke2018} also pointed out that the location of the planet can be shifted from the center of the gap.
In this paper, we further investigate effects of the planetary migration on the locations of the planet and gaps, which could be applied to the observation of the gas in near future.
From the recent ALMA observations, in the relatively outer region ($>30$~AU), the width of some observed gaps is relatively narrow and thus the planet mass estimated from the gap shape can be not so massive, typically Neptune size to sub-Jupiter size \citep[e.g.,][]{Dipierro_Price_Laibe_Hirsh_Cerioli_Lodato2015,Kanagawa2015b,Nomura_etal2016,Tsukagoshi2016,Zhang_DSHARP2018} when the gas viscosity is low as implied by observations \citep[e.g.][]{Pinte2016,Flaherty2015,Teague2016TWHydra}.
Hence, we focus on the observational signatures from the Neptune-sized planet in this paper.
We describe our model in Section~\ref{sec:basic_eq} and present our results in Section~\ref{sec:results}.
In Section~\ref{sec:discussion}, we discuss feasibility of observations and how to constrain the evolution of the planet from the observational signatures.
We summarize our results in Section~\ref{sec:summary}.

\section{Basic equations and our model description} \label{sec:basic_eq}
\subsection{Basic equations} \label{subsec:eq}
We investigate effects of a migrating planet on the gap structure, by carrying out two-dimensional hydrodynamic simulations with a planet.
In our simulations, we use a geometrically thin and non-self-gravitating disk.
We choose a two-dimensional cylindrical coordinate system $(R,\phi)$, and its origin locates at the position of the central star.
The velocity is denoted as $\vvec=(\vrad,\vphi)$, where $\vrad$ and $\vphi$ are the velocities in the radial and azimuthal directions.
The angular velocity is denoted by $\Omega=\vphi/R$.
We adopt a simple isothermal equation of state, in which the vertically integrated pressure $P$ is given by $c_s^2 \rhosurf$, where $c_s$ is the isothermal speed of sound.

The vertically integrated equation of continuity is
\begin{align}
	\dpar{\rhosurf}{t}+\nabla \cdot \left( \rhosurf \vvec \right) &=0.
	\label{eq:mass_eq}
\end{align}
The equations of motion are
\begin{align}
\dpar{\vvec}{t} + \vvec \cdot \nabla \vvec &= -\frac{\nabla P}{\rhosurf} - \nabla \Psi + \vec{f}_{\nu},
\label{eq:motion_eq}
\end{align}
where $\vec{f}_{\nu}$ represents the viscous force per unit mass \citep[c.f.,][]{Nelson_Papaloizou_Masset_Kley2000}.
The gravitational potential $\Psi$ is given by the sum of the gravitational potentials of the star and the planet as
\begin{align}
\Psi &= -\frac{\grav \mstar}{R} + \Psi_p + \frac{\grav \mpl}{\rp^2} R\cos\left(\phi-\phi_p \right),
\label{eq:gravpot}
\end{align}
where $\grav$ is the gravitational constant.
The first term of Equation~(\ref{eq:gravpot}) is the potential of the star and the third term represents the indirect terms due to planet--star gravitational interaction.
The second term is the gravitational potential of the planet, which is given by
\begin{align}
\Psi_{p} &= -\frac{\grav \mpl}{\left[ R^2+2R\rp \cos\left( \phi-\phi_{p} \right) + \rp^2 +\epsilon^2  \right]^{1/2}},
\label{eq:gravpot_planet}
\end{align}
where $\epsilon$ is a softening parameter.

\subsection{Our setup of hydrodynamic simulations}
To numerically solve Equations~(\ref{eq:mass_eq}) and (\ref{eq:motion_eq}), we use the two-dimensional numerical hydrodynamic code {\sc \tt FARGO}\footnote{See: \url{http://fargo.in2p3.fr/}} \cite{Masset2000}, which is an Eulerian polar grid code with a staggered mesh.
%Because of a fast advection algorithm that removes the azimuthally averaged velocity for the Courant time step \citep{Masset2000}.
The softening parameter $\epsilon$ in the gravitational potential of Equation~(\ref{eq:gravpot_planet}) is set to be $0.6$ times the disk scale height at the location of the planet.
Considering the existence of the circumplanetary disk, we exclude $60\%$ of the planets' Hill radius when calculating the force exerted by the disk on the planet.
For simplicity, we neglect the disk gas accretion onto the planet.

%Recent observations have revealed that in the outer region, the gap is relatively narrow.
%With reasonable values of $\alpha$ and $H_0$, the mass of the planet within the gap is estimated to be around the Neptune mass to sub-Jupiter mass \citep[e.g.][]{Kanagawa2015b,Zhang_DSHARP2018}.
From the relation between the planet mass and the width (and depth) of the gap, we can estimate the mass of the planet within the observed gap \citep[e.g.,][]{Kanagawa2015b,Rosotti_Juhasz_Booth_Clarke2016,Dong_Fung2017,Zhang_DSHARP2018}.
Recent observations have revealed relatively narrow gaps which can be carved by the planet around the Neptune-mass to sub-Jupiter mass, for instance, the gap at $\sim 70$~AU in the disk around HL~Tau ($\mpl\simeq 0.3M_J$ with $\alpha=10^{-3}$, the same $\alpha$ is assumed for the following planet mass) \citep[e.g.,][]{Kanagawa2016a,Jin_Li_Isella_Li_Ji2016}, the gap at $22$~AU in the disk of TW~Hya ($\mpl\simeq 0.06 M_J$) \citep{Tsukagoshi2016},  the gap at $97$~AU in the disk of RX J1615e ($\mpl=0.22M_J$) \citep{Dong_Fung2017}, and the gap at $69$~AU of Elias~27 disk ($\mpl \sim 0.1M_J$), the gap at $86$~AU in the disk of HD~163296 ($\mpl\simeq 0.3 M_J$) \citep{Zhang_DSHARP2018}.
Moreover, recent observation done by \cite{Tsukagoshi2019} has discovered the excess of the millimeter flux at $\sim 50$~AU.
From the excess of the flux, the mass of the planet is estimated as the Neptune size.
Hence, in this paper, we adopt the mass of the planet around Neptune-mass.

The recent observations give an upper limit on the $\alpha$-parameter on the viscosity for a few protoplanetary disks, namely, $\alpha \lesssim 10^{-3}$ in the disk of HD~163296 \citep{Flaherty2015,Flaherty2017} and in the disk of TW~Hya \citep{Teague2016TWHydra,Flaherty2018TWHydra}. 
Motivated by those observations, we adopt a relatively small value of $\alpha$.

The computational domain runs from $R=0.1R_0$ to $R=2.4R_0$, where we use a unit of the radius as an arbitrary value $R_0$ and a unit of the mass as $\mstar$ (the mass of the central star).
The domain is divided into 512 meshes in the radial direction (logarithmic equal spacing) and 1024 meshes in the azimuthal direction (equal spacing).
The orbital radius of the planet is initially set to be $R=R_0$.
The surface density is thus normalized by $\mstar/R_0^2$, and we choose $\mstar = 1M_{\odot}$ as the fiducial value.
Since focusing on the planet orbiting at larger radii, we assume $R_0=100$~AU in this paper.
For convenience, we define $t_0$ as the Keplerian orbital time at $R=R_0$.
We compute the migration of the planet during $t=1000\ t_0$ (which corresponds to 1~Myr when $R_0=100$~AU).
We assume a uniform distribution of the disk aspect ratio, $h/R=0.05$.
Since we adopt a structure in steady state of viscous accretion disk \citep[c.f.][]{Lynden-Bell_Pringle1974} as the initial condition, the unperturbed distribution of the surface density given by
\begin{align}
\sigmaun(R) = \Sigma_0 \bracketfunc{R}{R_0}{-1/2}.
\label{eq:sigma_un}
\end{align}
The initial angular velocity of the gas is given by $\omegak \sqrt{1-2\eta}$, where $\eta = (1/2)(\h/R)^2 d\ln P/d\ln R$.
The initial radial drift velocity of the gas is given by $v_R = -3\nu/(2R)$.
The parameters we investigate in this paper are summarized in Table~\ref{tab:params}.
Note that when $\mstar=1M_{\odot}$ and $R_0=100$~AU, the surface density is $0.9\rm{g/cm}^2 (\Sigma_0/10^{-3})$ at $R=100$~AU.
For simplicity we do not consider growth of the mass of the planet, whereas the planetary orbit varies with time according to the disk-planet interaction.
\begin{deluxetable}{cccc}
\tablenum{1}
\tablecaption{Parameters\label{tab:params}}
\tablewidth{0pt}
\tablehead{
\colhead{$\alpha$} & \colhead{$\Sigma_0$} & \colhead{$\mpl/\mstar$} & \colhead{$H_0$} 
}
\startdata
$5 \times 10^{-5}$ & $[1,3,5,7,10] \times 10^{-4}$ & $5\times 10^{-5}$ & $0.05$\\
$1 \times 10^{-4}$ & $[1,3,5,7,10] \times 10^{-4}$ & $5\times 10^{-5}$ & $0.05$\\
$3 \times 10^{-4}$ & $[1,3,5,7,10] \times 10^{-4}$ & $5\times 10^{-5}$ & $0.05$\\
$5 \times 10^{-4}$ & $[1,5,10] \times 10^{-4}$ & $5\times 10^{-5}$ & $0.05$\\
$1 \times 10^{-3}$ & $[1,3,5,7,10] \times 10^{-4}$ & $5\times 10^{-5}$ & $0.05$\\
$1 \times 10^{-4}$ & $[1,3,5,7,10] \times 10^{-4}$ & $1\times 10^{-4}$ & $0.05$\\
$1 \times 10^{-3}$ & $[1,3,5,7,10] \times 10^{-4}$ & $1\times 10^{-4}$ & $0.05$\\
\hline
$5 \times 10^{-5}$ & $[1,5,10] \times 10^{-4}$ & $5\times 10^{-5}$ & $0.07$\\
$1 \times 10^{-4}$ & $[1,5,10] \times 10^{-4}$ & $5\times 10^{-5}$ & $0.07$\\
$1 \times 10^{-4}$ & $[1,5,10] \times 10^{-4}$ & $1\times 10^{-4}$ & $0.07$\\
\hline
$5 \times 10^{-5}$ & $[1,5,10] \times 10^{-4}$ & $1\times 10^{-4}$ & $0.1$\\
\enddata
%\tablecomments{}
\end{deluxetable}

\REDD{At the inner and outer boundaries, the velocity of the gas is set to be the initial value.}
The surface density of the gas is also set so that the mass flux is constant.
We define wave-killing zones which are located from $R_{\rm out}-0.1R_0$ to $R_{\rm out}$ for the outer boundary and from $R_{\rm in}$ to $R_{\rm in}+0.1R_0$ for the inner boundary, where $R_{\rm out}$ and $R_{\rm in}$ are the radius of the outer and inner boundaries, respectively.
To avoid an artificial wave reflection, we force all the physical quantities to be azimuthally constant within the wave-killing zones, by overwriting the quantities with their azimuthal average at every time step \citep[c.f.,][]{Val-Borro_etal2006, Kanagawa_Ueda_Muto_Okuzumi2017}.

\section{Results of hydrodynamic simulations} \label{sec:results}
\subsection{Radial shift between the locations of the planet and the gap} \label{subsec:radial_shift}
Here, we define the location of the gap as the location where the azimuthally averaged surface density normalized by the unperturbed surface density $\sigmaun$ is the minimum in the region of $R>\rp$ \footnote{Since the planet migrates only inward in our simulations, the minimum surface density related to the gap always lies on the outer disk of the planet.}.
We denote this radius of the gap as $\rgap$, and $\rp$ denotes the orbital radius of the planet.
In our simulations, the secondary gap is formed in the inner disk of the planet, as shown by e.g., \cite{Bae_Zhu_Hartmann2017,Dong_Li_Chiang_Li2017}.
For convenience, we define the location of the secondary gap $\rgapsecond$ as the position where $\rhosurf/\sigmaun$ takes the first local minimum from $\rp$ in the inner disk.

\begin{figure}
\plotone{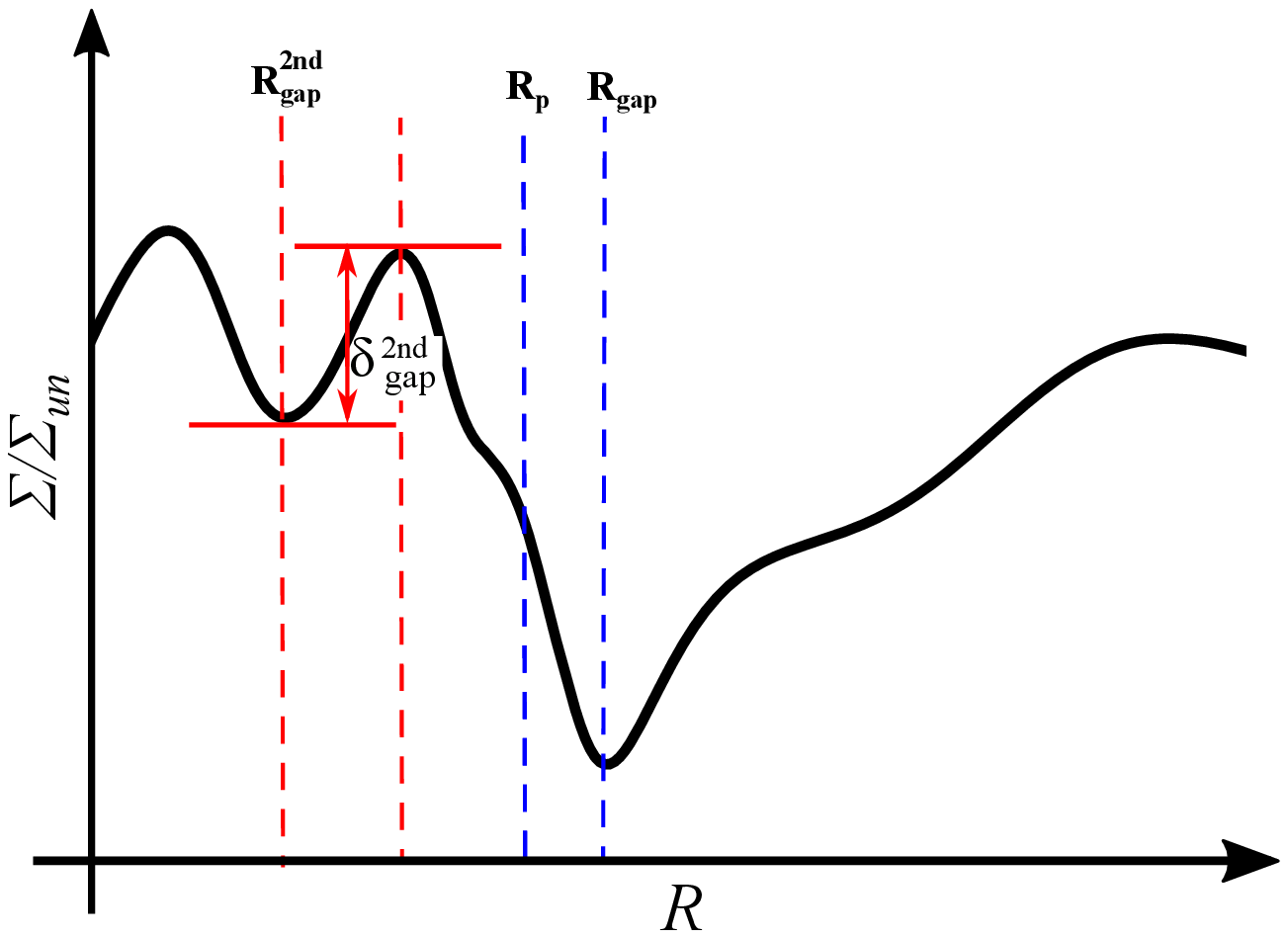}
\caption{
Schematic picture for the definition of the location of the (primary) gap $\rgap$, the location of the secondary gap $\rgapsecond$, and the depth of the secondary gap $\depsecond$.
\label{fig:defs}}
\end{figure}

The depth of the secondary gap $\depsecond$ is defined by the ratio of the surface densities at $R=\rgapsecond$ and the position where $\rhosurf/\sigmaun$ takes the first local maximum from $\rp$ in the inner disk\footnote{To avoid the structure in the vicinity of the planet, we exclude the region between $\rp$ and $\rp-1.5\max(\rhill,\hp)$ when searching the local maximum of the surface density, where $\rhill$ denotes the Hill radius of the planet, $\rhill=\rp[\mpl/(3\mstar)]^{1/3}$ and $\hp$ is a disk scale height at $\rp$}.
In Figure~\ref{fig:defs}, we illustrate the definitions of $\rgap$, $\rgapsecond$ and $\depsecond$.

First we show the results in the case of $\mpl/\mstar=5\times 10^{-5}$, $h/R=0.05$, and $\alpha=1\times 10^{-4}$, as the fiducial case.
\begin{figure*}
\plottwo{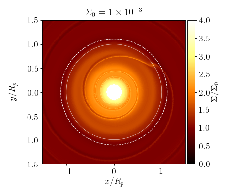}{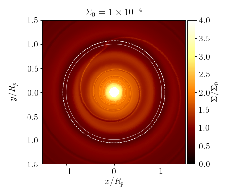}
\caption{
Two-dimensional distributions of the gas surface density in the cases of $\Sigma_0=1\times 10^{-3}$ (Left) and $\Sigma_0=10^{-4}$ (Right) at $t=1000\ t_0$.
The mass of the planet is $\mpl/\mstar=5\times 10^{-5}$, and $H_0=0.05$ and $\alpha=10^{-4}$, respectively.
The white solid and dashed lines denote the orbital radius of the planet ($R=\rp$)and the radial position of the gap (location with the minimum surface density within the gap, $R=\rgap$).
The vertical and horizontal axes are normalized by $\rp$, and $\rp=0.61R_0$ in the right panel and $\rp=0.95R_0$ in the left panel, respectively (see also Figure~\ref{fig:Evo_ap_q5e-5_a1e-4_h5e-2}).
\label{fig:2Ddist_q5e-5_h5e-2_a1e-4}}
\end{figure*}
\begin{figure}
\plotone{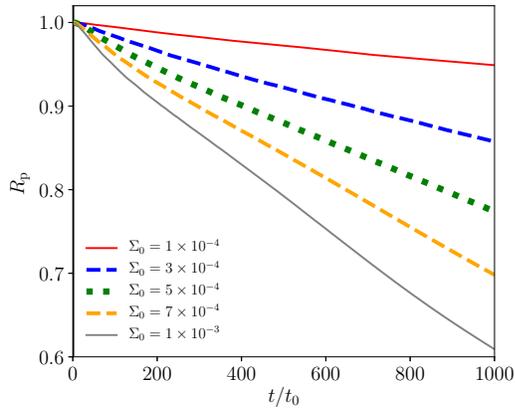}
\caption{
Time variations of the semi-major axis of the planet for various values of $\Sigma_0$.
The planet mass, aspect ratio and the viscosity are the same as those of the case shown in Figure~\ref{fig:2Ddist_q5e-5_h5e-2_a1e-4} ($\mpl/\mstar=5\times 10^{-5}, H_0=0.05, \alpha=10^{-4}$).
\label{fig:Evo_ap_q5e-5_a1e-4_h5e-2}}
\end{figure}
Figure~\ref{fig:2Ddist_q5e-5_h5e-2_a1e-4} illustrates the two-dimensional distributions of the gas surface density at $t=1000\ t_0$ in the fiducial cases with $\Sigma_0=10^{-3}$ and $\Sigma_0=10^{-4}$.
In both the cases, the planet migrates inward and the inward migration velocity is faster with the larger $\Sigma_0$ (Figure~\ref{fig:Evo_ap_q5e-5_a1e-4_h5e-2}).
As can be seen from Figure~\ref{fig:2Ddist_q5e-5_h5e-2_a1e-4}, the orbital radius of the planet ($\rp$, it is denoted by the white solid cycle in the figure) and the radius of the gap ($\rgap$, the white dashed cycle) are different in both the cases.
However, the radial difference between $\rgap$ and $\rp$ in the case with $\Sigma_0=10^{-3}$ is larger than that in the case of $\Sigma_0=10^{-4}$.
This radial shift between $\rp$ and $\rgap$ is also pointed out by \cite{Meru_Rosotti_Booth_Nazari_Clarke2018}.
As shown by \cite{Bae_Zhu_Hartmann2017} and \cite{Dong_Li_Chiang_Li2017}, moreover, the secondary gap is formed at the inner disk of the planet, since we assume the small value of the $\alpha$ parameter.
Note that we carried out hydrodynamic simulations with a higher resolution (1024 meshes in radial direction and 2048 meshes in azimuthal direction) and confirmed that a migration velocity and a distribution of azimuthal averaged surface density are converged (see Appendix~\ref{sec:reso_dep}).

\begin{figure}
\plotone{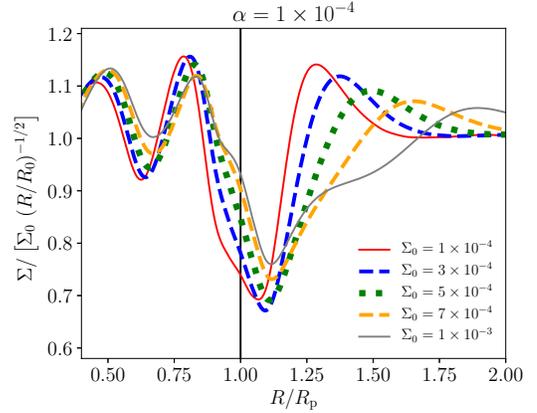}
\caption{
Azimuthally averaged surface density normalized by the initial surface density ($=\Sigma_0 (R/R_0)^{-1/2}$) for the cases of $\alpha=10^{-4}$ at $t=1000\ t_0$.
The planet mass and aspect ratio are the same as those of the case shown in Figure~\ref{fig:2Ddist_q5e-5_h5e-2_a1e-4}.
\label{fig:1D_norm_dist_q5e-5_a1e-4}}
\end{figure}
Dependence of the radial difference between $\rgap$ and $\rp$ is clearly shown in Figure~\ref{fig:1D_norm_dist_q5e-5_a1e-4} which illustrates the azimuthally averaged surface density normalized by $R^{-1/2}$ for various values of $\Sigma_0$.
The planet mass, aspect ratio, and the viscosity are the same as those in the case shown in Figure~\ref{fig:2Ddist_q5e-5_h5e-2_a1e-4}.
The location with the smallest $\Sigma/\sigmaun$ in the outer disk of the planet corresponds to $\rgap$ (see Figure~\ref{fig:defs}).
As can be seen from Figure~\ref{fig:1D_norm_dist_q5e-5_a1e-4}, the difference between $\rgap$ and $\rp$ becomes larger, with the larger $\Sigma_0$.
When $\Sigma_0=10^{-4}$, the planet locates close to the location of the gap bottom.
On the other hand, the planet locates at the inner edge of the gap and the $\rp$ and $\rgap$ are significantly different from each other when $\Sigma_0=10^{-3}$.

In all the cases shown in Figure~\ref{fig:1D_norm_dist_q5e-5_a1e-4}, a visible secondary gap is formed.
The location of the secondary gap weakly depends on $\Sigma_0$, namely, it forms at slightly smaller radii with a smaller $\Sigma_0$ ($R^{\rm 2nd}_{\rm gap}/\rp = 0.67$ in the case with $\Sigma_0=10^{-3}$, and $R^{\rm 2nd}_{\rm gap}/\rp = 0.63$ in the case with $\Sigma_0=10^{-4}$).
The depths of the gap in the vicinity of the planet (primary gap) and the secondary gap hardly depends on $\Sigma_0$.

The radial shift between the locations of the planet and the gap depends on the $\alpha$ parameter and the aspect ratio.
\begin{figure}
%\plotone{NormAvgDens_q5e-5_a1e-3.eps}
%\plotone{NormAvgDens_q5e-5_a1e-4_h0.07.eps}
\plotone{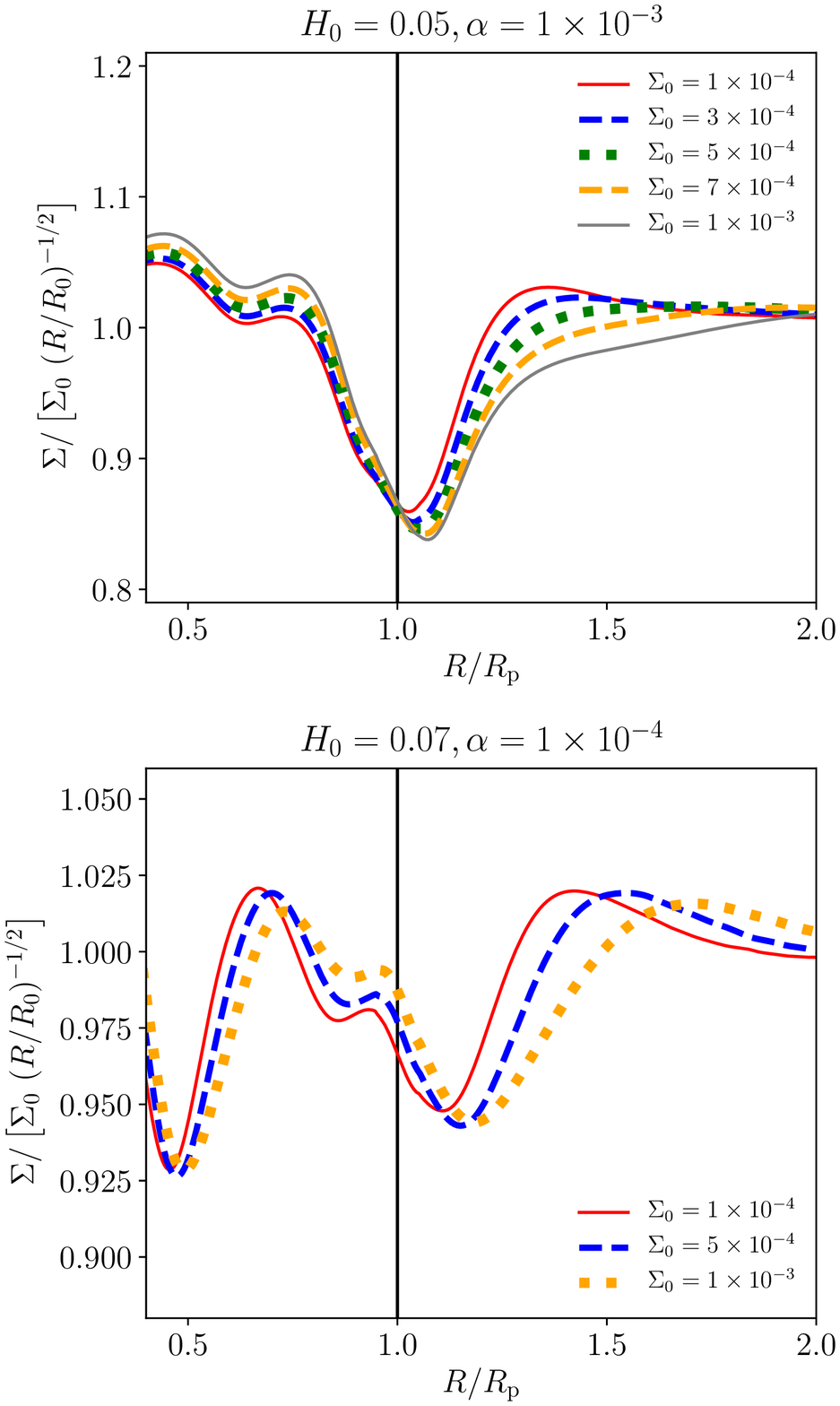}
\caption{
The same as Figure~\ref{fig:1D_norm_dist_q5e-5_a1e-4}, but for the case with $\alpha=10^{-3}$ (upper panel) and for the case with $H_0=0.07$.
\label{fig:1D_norm_dist_q5e-5_a1e-3_and_h0.07}}
\end{figure}
The upper panel of Figure~\ref{fig:1D_norm_dist_q5e-5_a1e-3_and_h0.07} shows the azimuthally averaged surface density normalized by the initial surface density distribution in the case with $\alpha=10^{-3}$, and the planet mass and aspect ratio are the same as those in the fiducial case.
As shown in Figure~\ref{fig:1D_norm_dist_q5e-5_a1e-3_and_h0.07}, the radial difference hardly depends on $\Sigma_0$ when $\alpha=10^{-3}$.
Although the secondary gap is formed in this case at $R/\rp = 0.64$, moreover, it is not visible because its depth is very shallow.
The lower panel of Figure~\ref{fig:1D_norm_dist_q5e-5_a1e-3_and_h0.07} shows the case with $H_0=0.07$.
Even when the aspect ratio is larger than that in the fiducial case, the radial difference between $\rp$ and $\rgap$ becomes larger with the larger $\Sigma_0$, which is the same as that shown in Figure~\ref{fig:1D_norm_dist_q5e-5_a1e-4}.
Although the gap around the planet (primary gap) is shallower than that in the case of Figure~\ref{fig:1D_norm_dist_q5e-5_a1e-4} due to the larger $H_0$, the depth of the secondary gap is similar to that in the case shown in Figure~\ref{fig:1D_norm_dist_q5e-5_a1e-4}.
The location of the secondary gap is formed at the smaller radii ($R/\rp \simeq 0.5$) than that in the case shown in Figure~\ref{fig:1D_norm_dist_q5e-5_a1e-4}.
In Section~\ref{subsec:2ndgap}, we discuss the parameter dependence of the secondary gap.

\subsection{Empirical formula for the radial shift between the locations of the planet and the gap} \label{subsec:empirical_lformula}
\subsubsection{Timescales}
As shown in the previous subsections, the radial difference between the locations of the planet and the gap becomes larger with the lower viscosity and higher surface density.
This radial difference may be explained by the ratio of the timescales of the gap formation and the radial migration.
According to \cite{Kanagawa2017b}, the timescale of the gap formation $\tgap$ is 
\begin{align}
\tgap &= \left(\frac{\Delta_{\rm gap}}{2\rp} \right)^2  \bracketfunc{\hp}{\rp}{-2} \alpha^{-1} \omegakp^{-1},\label{eq:tgap}
\end{align}
where $\Delta_{\rm gap}$ is the half width of the gap which can be given by
\begin{align}
\frac{\Delta_{\rm gap}}{\rp} &= 0.41 K'^{1/4}, \label{eq:delta_gap}\\
K'&= \bracketfunc{\mpl}{\mstar}{2} \bracketfunc{\hp}{\rp}{-3} \alpha^{-1} \label{eq:kprime},
\end{align}
where $\omegak$ denote the Keperian angular velocity, and the subscription 'p' indicates the value at $R=\rp$.
%Alternatively, $\tgap$ is expressed by
Equation~(\ref{eq:tgap}) can be rewritten as 
\begin{align}
\tgap&=4.24 \bracketfunc{\mpl/\mstar}{5\times 10^{-5}}{} \bracketfunc{\hp/\rp}{0.05}{-7/2} \nonumber\\
& \quad \times \bracketfunc{\alpha}{10^{-4}}{-3/2} \bracketfunc{\mstar}{1 M_{\odot}}{-1/2} \bracketfunc{\rp}{50\ \mbox{AU}}{3/2} \ \mbox{Myr}
\label{eq:tgap_wdim}
\end{align}
When the radial migration is progressing slower than its gap formation, the gap shape can reach that in steady state before the planet moves significantly.
In this case, the migration timescale $\tmig{}_{\rm ,steady}$ can be given by \citep{Kanagawa_Tanaka_Szuszkiewicz2018},
\begin{align}
\tmig{}_{\rm ,steady} &= \frac{\sigmaunp}{\sigmagap} \tau_{I},\nonumber \\
&=\left(1+0.04K \right) \tau_{I} \label{eq:tmig}
\end{align}
where $\sigmaunp$ is the unperturbed surface density at $\rp$ and $K$ is defined by
\begin{align}
K&=\bracketfunc{\mpl}{\mstar}{2}\bracketfunc{\hp}{\rp}{-5}\alpha^{-1},
\label{eq:k}
\end{align}
and $\sigmagap$ is the surface density at the bottom of the gap in steady state:
\begin{align}
\frac{\sigmagap}{\sigmaunp} &= \frac{1}{1+0.04K}.
\label{eq:sigmagap}
\end{align}
The migration timescale predicted by the type I migration $\tau_{I}$ is expressed by
\citep{Tanaka_Takeuchi_Ward2002,Paardekooper_Baruteau_Crida_Kley2010}
\begin{align}
\tau_{I} &= \frac{1}{2c} \bracketfunc{\mpl}{\mstar}{-1}\bracketfunc{\mstar}{\sigmaunp \rp^2}{} \bracketfunc{\hp}{\rp}{2} \omegakp^{-1},\label{eq:tau_I}\\
&= 1.68 \bracketfunc{c}{3}{} \bracketfunc{\mpl/\mstar}{5\times 10^{-5}}{-1} \bracketfunc{\sigmaunp}{1\mbox{ g/cm}^2}{-1} \nonumber \\
&\qquad \bracketfunc{\hp/\rp}{0.05}{2} \bracketfunc{\mstar}{1 M_{\odot}}{1/2} \bracketfunc{\rp}{50\mbox{ AU}}{1/2} \mbox{ Myr} \label{eq:tau_I_wdim},
\end{align}
where the coefficient $c$ is related to the radial distributions of $\rhosurf$ and $\h$, here we adopt $c=3$.
When $\tgap$ is much longer than $1000\ t_0$, the migration timescale can not reach the value in steady state, which is given by Equation~(\ref{eq:tmig}).

When the viscosity is very low as in the case we assume in this paper, the gap-opening time is much longer than the migration timescale given by Equation~(\ref{eq:tmig}).
In this case, the planet migrates with a gap which is not fully formed and the migration timescale must be shorter than that given by Equation~(\ref{eq:tmig}), due to the incomplete gap formation.
Considering this effect of the incomplete formation of the gap, \cite{Kanagawa_Tanaka_Szuszkiewicz2018} also gives the following formula:
\begin{align}
\tmig &= \left[1+0.04K\left(1-e^{-t/\tgap}\right)\right] \tau_{I}.
\label{eq:tmig_incomp}
\end{align}
In particular, the migration timescale is approximately given by $\tau_{I}$ during $1000 t_0$ (our simulation time) when the viscosity is very small, namely $\alpha \sim 1\times 10^{-4}$, since $\tgap \gg 1000 t_0$.

\subsubsection{Empirical formula}
\begin{figure}
\plotone{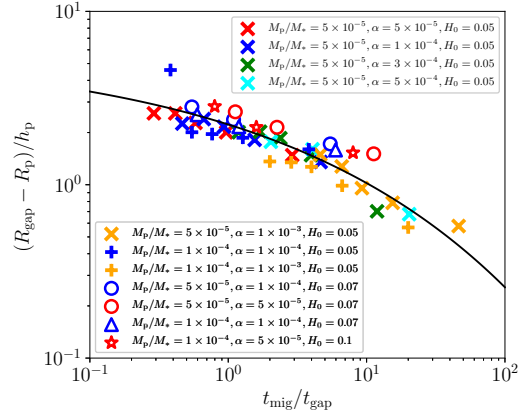}
\caption{
Radial shift ($\Delta$) between the planetary position ($\rp$) to the location of the gap ($\rgap$) for various planet masses, $\alpha$ and $\Sigma_0$. 
The thin solid line denotes our empirical formula (Equation~\ref{eq:empirical_fit}).
\label{fig:Shift_vs_timeratio}}
\end{figure}
We estimate the gap-opening timescale $\tgap$ by Equation~(\ref{eq:tgap}) and the migration timescale $\tmig$ by Equation~(\ref{eq:tmig_incomp}).
Figure~\ref{fig:Shift_vs_timeratio} shows the radial difference between the locations of the planet and the gap, as a function of the ratio of $\tmig$ and $\tgap$.
As can be seen from the figure, the shift can be fitted by the following empirical formula as a function of $\tmig/\tgap$, as
\begin{align}
\frac{\rgap-\rp}{\hp} &= 6.05 \exp\left[-\bracketfunc{\tmig}{\tgap}{0.25}\right].
\label{eq:empirical_fit}
\end{align} 

\begin{figure}
\plotone{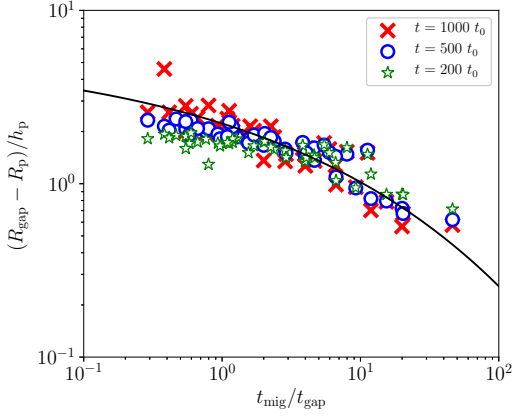}
\caption{
The same as Figure~\ref{fig:Shift_vs_timeratio}, but for different moments, $t=200\ t_0$ (0.2~Myr when $R_0=100$~AU), $t= 500 t_0$ (0.5~Myr) and $t=1000 t_0$ (1~Myr shown in Figure~\ref{fig:Shift_vs_timeratio}).
\label{fig:Shift_vs_timeratio_timevar}}
\end{figure}
In Figure~\ref{fig:Shift_vs_timeratio_timevar}, we show the relation between the radial difference between $\rp$ and $\rgap$ at the different moments.
As can be seen from the figure, the scaling relation given by Equation~(\ref{eq:empirical_fit}) is still valid, regardless of the time.
This result implies that Equation~(\ref{eq:empirical_fit}) can be applied to protoplanetary disks, regardless of the evolution phases, from Class I to Class II.
Observational implications of Equation~(\ref{eq:empirical_fit}) is discussed in Section~\ref{subsec:obs_implications}.

\subsection{Secondary gap} \label{subsec:2ndgap}
When the viscosity is relatively low, the planet can form the secondary gap in the inner disk of the planet \citep[e.g.,][]{Bae_Zhu_Hartmann2017,Dong_Li_Chiang_Li2017}.
From the depth and the location of the secondary gap, we can constrain the viscosity and the scale height.
\begin{figure}
\plotone{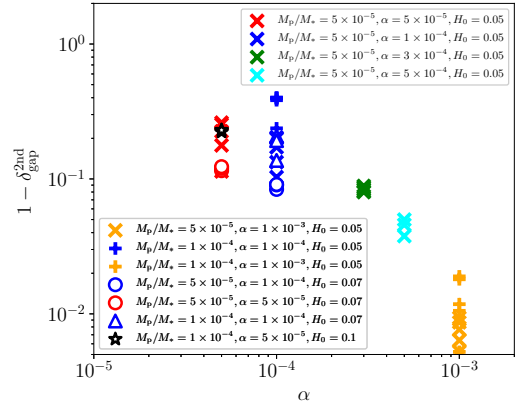}
\caption{
Depths of the secondary gap for various runs at $t=1000\ t_0$.
\label{fig:2ndGap_depth}}
\end{figure}
In Figure~\ref{fig:2ndGap_depth}, we illustrate the depth of the secondary gaps given by our simulations ($\depsecond$) at $t=1000 \ t_0$.
When the $\alpha$-parameter is relatively large as $\sim 10^{-3}$, only the shallow gap is formed.
In this case, the secondary gap could not be observed.
On the other hand, in the case with the low viscosity, namely $\alpha \lesssim 3\times 10^{-4}$, the relatively deep gap is formed, namely $\depsecond \lesssim 0.9$.
For $\alpha\lesssim 3\times 10^{-4}$, the depth of the secondary gap is not sensitive to $\alpha$.
Hence, we can obtain the constraint of $\alpha \lesssim 3 \times 10^{-4}$ if the secondary gap is observed.
Otherwise, we can constrain the lower limit as $\alpha \gtrsim 3 \times 10^{-4}$.

\begin{figure}
\plotone{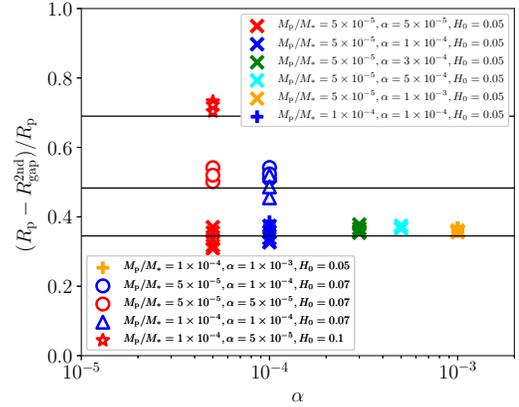}
\caption{
Locations of the secondary gap at $t=1000 \ t_0$.
The solid lines indicate Equation~(\ref{eq:location_2ndgap}) with $(h/R)^{\rm 2nd}_{\rm gap}=0.05$, $0.07$, and  $0.1$ from the bottom.
Note that in the cases shown in this figure, the aspect ratio is constant throughout the computational domain.
\label{fig:2ndGap_location}}
\end{figure}
\begin{figure}
\plotone{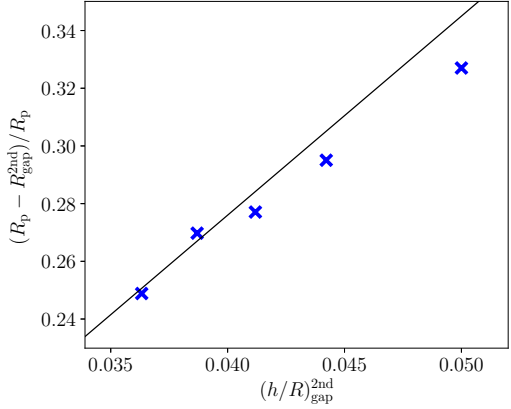}
\caption{
Location of the secondary gap at $t=1000 \ t_0$ as a function of the aspect ratios at the secondary gap when the aspect ratio depends on radii, $h/R = H_0 (R/R_0)^{f}$.
From the left, the crosses correspond to the cases of $f=0.5$, $f=0.35$, $f=0.25$, $f=0.15$, and $f=0$, respectively.
The planet mass, the value of $\alpha$-parameter, the disk aspect ratio at $R=R_0$ are the same as those in our fiducial case.
The solid line denotes Equation~(\ref{eq:location_2ndgap}).
\label{fig:2ndGap_location_vs_findex}}
\end{figure}
Figure~\ref{fig:2ndGap_location} illustrates the locations of the secondary gap.
As different from the dependence of $\depsecond$, the location does not depend on $\alpha$.
We found that $\rgapsecond - \rp$ is proportional to $H_0$.
The location of the secondary gap also depends on a radial distribution of the aspect ratio.
To investigate effects of the radial distribution of the aspect ratio, assuming the $h/R = H_0 (R/R_0)^{f}$, we carried out additional hydrodynamic simulations with different values of $f$ ($f=0.15,0.25,0.35,0.5$).
The other parameters (e.g., $\mpl,\alpha,H_0$) are the same as those of our fiducial case. 
Results of these simulations are shown in Figure~\ref{fig:2ndGap_location_vs_findex}.
As can be seen from the figure, the location of the secondary gap is proportional to the aspect ratio at the secondary gap $(h/R)_{\rm gap}^{\rm 2nd}$, rather than that at the planetary orbital radius.

Taking into account Figures~\ref{fig:2ndGap_location} and \ref{fig:2ndGap_location_vs_findex}, we can obtain the relation between the location of the secondary gap and the aspect ratio as
\begin{align}
\frac{\rgapsecond-\rp}{\rp} = 0.345 \bracketfuncb{(h/R)^{\rm 2nd}_{\rm gap}}{0.05}{}.
\label{eq:location_2ndgap}
\end{align}
%As can be seen from Figures~\ref{fig:2ndGap_location} and \ref{fig:2ndGap_location_vs_findex}, the location of the secondary gap can be reasonably scaled by Equation~(\ref{eq:location_2ndgap}).
\begin{figure}
\plotone{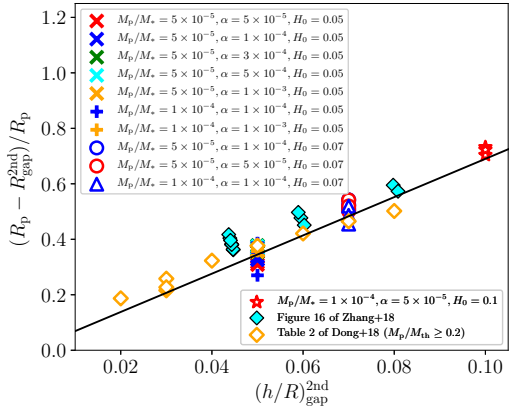}
\caption{
The same as Figure~\ref{fig:2ndGap_location}, but it is shown by a function of the aspect ratio at the location of the secondary gap.
The solid line denotes Equation~(\ref{eq:location_2ndgap})
\label{fig:2ndGap_location_vs_aspectr}}
\end{figure}
Figure~\ref{fig:2ndGap_location_vs_aspectr} shows the same as that shown in Figure~\ref{fig:2ndGap_location}, but as a function of the aspect ratio at the location of the secondary gap.
This figure shows that the fitting formula of Equation~(\ref{eq:location_2ndgap}) also can well reproduce the location of the secondary gap.
We should note that the position of the secondary gap also depends on the migration speed, though its dependence is weak as pointed out in Section~\ref{subsec:radial_shift}.
When the migration is slow in the case with a smaller $\Sigma_0$, $1-\rgapsecond/\rp$ is slightly larger.
Because of it, the locations of the secondary gap are spread within the range of $\sim 0.1 \rp$ in Figure~\ref{fig:2ndGap_location_vs_aspectr} even when the mass of the planet, aspect ratio and viscosity are the same. 
We also confirmed this trend by carrying out the simulation with a fixed orbit.
In the case of the planet with a fixed orbit, the value of $1-\rgapsecond/\rp$ is almost the same as the upper values shown in Figure~\ref{fig:2ndGap_location_vs_aspectr}.

\cite{Dong_Li_Chiang_Li2018} and \cite{Zhang_DSHARP2018} have also shown that the location of the secondary gap depends on the disk scale height, and they give similar scaling relations to Equation~(\ref{eq:location_2ndgap}).
In Figure~\ref{fig:2ndGap_location_vs_aspectr}, we plot the data extracted from Table~2 of \cite{Dong_Li_Chiang_Li2018} (data for $\mpl/M_{\rm th}\geq 0.2$, where $M_{\rm th}=(\hp/\rp)^3$) and Figure~16 of \cite{Zhang_DSHARP2018}.
Equation~(\ref{eq:location_2ndgap}) also can reproduce both the results given by \cite{Dong_Li_Chiang_Li2018} and \cite{Zhang_DSHARP2018}.
In the scaling of \cite{Zhang_DSHARP2018}, the power of $h/R$ is smaller than that of Equation~(\ref{eq:location_2ndgap}), namely $\rp-\rgapsecond \propto (h/R)^{0.58}$, though it hardly depends on the mass of the planet similar to Equation~(\ref{eq:location_2ndgap}).
The difference of the power of $h/R$ could be caused by the spatial distribution of the aspect ratio.
Since \cite{Zhang_DSHARP2018} assumes $h/R\propto R^{0.25}$, the aspect ratio at the location of the secondary gap is smaller than that at the location of the planet.

\cite{Dong_Li_Chiang_Li2018} assume a constant aspect ratio ($h/R= \mbox{const}$) and gives a similar power of $h/R$, namely $\propto (h/R)^{1.3}$, but it depends on $\mpl^{0.2}$.
The formula given by \cite{Dong_Li_Chiang_Li2018} give a better fit when the planet is small, namely $\mpl/M_{\rm th} \lesssim 0.2$.
However, the prediction given by the formula of \cite{Dong_Li_Chiang_Li2018} does not fit the results of simulations when $\mpl/M_{\rm th} \gg 1$ (see Appendix~\ref{sec:location_2ndgap}).
For a relatively large planet which can be detected by ALMA, Equation~(\ref{eq:location_2ndgap}) may be convenient, rather than the formula given by \cite{Dong_Li_Chiang_Li2018}.

\section{Discussion} \label{sec:discussion}
\subsection{Feasibility of observations} \label{subsec:observability}
In the above section, we show that when the inward migration of the planet is faster than the gap-opening, the orbital radius of the planet $\rp$ is smaller than that of the gap $\rgap$.
If such a difference is observed, it could be evidence that the planet is formed in the outer region and it is migrating inward quickly.
We also found the scaling relation of Equation~(\ref{eq:empirical_fit}) which gives the relation of the radial difference between $\rp$ and $\rgap$ and the ratio of the timescale of the migration and gap-opening given by Equations~(\ref{eq:tmig_incomp}) and (\ref{eq:tgap}), respectively.
If the secondary gap is observed, we also can constrain the disk viscosity and aspect ratio as shown in Section~\ref{subsec:2ndgap}.
In this subsection, we discuss feasibility of the observations of gap profile of gas and excess of the dust emission from a planet embedded within the disk.

The CO line emission has been detected by the observation with ALMA in Band~7 in Cycle~2 at the disk around TW~Hya \citep{Nomura_etal2016}.
By using the line emission from \hco and \cho J= 3--2, \cite{Nomura_etal2016} has obtained the column density distribution of CO.
Since the \cho emission is likely to be optically thin in an outer region of the disk, the CO column density can be directly compared with the gas surface density given by hydrodynamic simulations \footnote{Strictly speaking, the CO emission comes from a location where is slightly above the midplane, because most of the CO molecules are frozen out on the surface o the dust at the midplane. Because of it, we may underestimate the absolute value of the gas density. However, the CO density estimated from the CO emission could be proportional to the gas density. Hence we could know the shape of the gap from the CO emission.}.
In the recent observation with higher angular resolution ($\sim$~0.15~arcsec, $\sim 9$~AU resolution) and 2.3 hours on-source integration time, the gap profile of CO is possibly detected around $\sim 50$~AU (Nomura et al. in prep).
With ALMA in Band~7 in Cycle~3, \cite{Tsukagoshi2019} has detected the point source in dust continuum emission at $52$~AU in the disk around TW~Hya.
The angular resolution and the on-source integration time of that observation are $\sim$~0.043~arcsec ($\sim 3$~AU resolution) and 3.5~hours, respectively.

In the basis of the observations of TW~Hya mentioned above, we estimate feasibility of the detection of CO and dust point source in other protoplanetary disks.
We assume that a distance to the protoplanetary disk is around $130$pc in this estimate (the distance to TW~Hya is about $60$pc).
Because of the larger distance, it takes higher angular resolution and longer integration time to detect the CO line and a point source in dust emission.
To achieve the same spatial resolutions of \cite{Nomura_etal2016} and \cite{Tsukagoshi2019}, the angular resolution of $0.07$~arcsec for CO line observation and $0.023$~arcsec for the dust continuum observation are required.
Since a required integration time is proportional to 4th power of the angular resolution, it can be estimated as $\sim 50$~ hours, which is unreasonably long at the current moment.
However, since the gap width is scaled by the orbital radius of the planet \citep[see e.g.,][]{Kanagawa2016a}, the gap profile can be detected in an outer region with lower angular resolution.
When $\rp \sim 100$~AU, one could observe the gap of CO emission and the point source of the dust emission with the similar angular resolutions and integration times, for the disk with the distance of $\simeq 130$pc.

In the outer region ($\sim 100$~AU), the CO column density and the temperature may be smaller and lower than these of TW~Hya at $\sim 50$~AU. 
With lower CO column density and temperature, a longer integration time would be required due to weak emission.
For instance, however, \cite{Isella2016} has shown that in the case of the disk around HD~163296, the \cho J = 2--1 emission at $100$~AU ($\sim 0.65$~Jy/arcsec$^2$/(km/s)) is comparable to the \cho J= 3--2 emission at $50$~AU in the disk of TW~Hya ($\sim 0.14$~Jy/arcsec$^2$/(km/s)).
Moreover, DSHARP program has observed the $^{12}$~CO J= 2--1 emission and revealed that in some disks, i.e., AS~209 \citep{Guzman_DSHARP_AS209} and HD~143006 \citep{Perez_DSHARP_HD143006}, the $^{12}$~CO emission around 100~AU is comparable with or larger than the \cho emission at 50~AU in TW~Hya disk.
For the disks around Herbig stars, the CO emission at $100$~AU can be comparable with that at $50$~AU in the disk around TW~Hya.
The gap in the CO emission and the point source in the dust emission could be observed with the similar angular resolutions and integration times as these of \cite{Nomura_etal2016} and \cite{Tsukagoshi2019}. 

Detecting the secondary gap in the gas might be challenging because it is shallow and narrow, moreover formed in an inner region than the primary gap.
However, the secondary gap is easier to be observed by the observations of the dust continuum.
We can estimate the disk scale height from the location of the secondary gap measured by the location of the secondary gap by dust observations, by Equation~(\ref{eq:location_2ndgap}).
The depth of the secondary gap could be affected by the size of the dust grains, as well as the gas viscosity.
Hence, we need to take care of the size of the dust to estimate the upper/lower limit of viscosity.

\subsection{Observational implications} \label{subsec:obs_implications}
In this \RED{subsection}, we discuss what can constrained when the gap and planet are observed.
The difference between $\rp$ and $\rgap$ depends on $\tmig/\tgap$, that is, it depends on the mass of the planet, the disk viscosity, aspect ratio and the gas surface density of the disk, as can be seen from Equation~(\ref{eq:empirical_fit}).
The mass of the planet can be estimated from the excess of the flux at the planet location \citep[e.g.,][]{Ayliffe_Bate2009,Wang_Bu_Shang_Gu2014,Szulagyi_Cilibrasi_Mayer2018}, and the aspect ratio can be also estimated from the brightness temperature of the dust emission if the dust emission is detected \citep[e.g.,][]{Nomura_etal2016}.
On the other hand, the viscosity and the gas surface density (not dust and CO densities) are relatively difficult to be constrained from the observation.
However, by using Equation~(\ref{eq:empirical_fit}), we can constrain the viscosity and the gas surface density from the observed radial difference between $\rp$ and $\rgap$.

\begin{figure*}
\plottwo{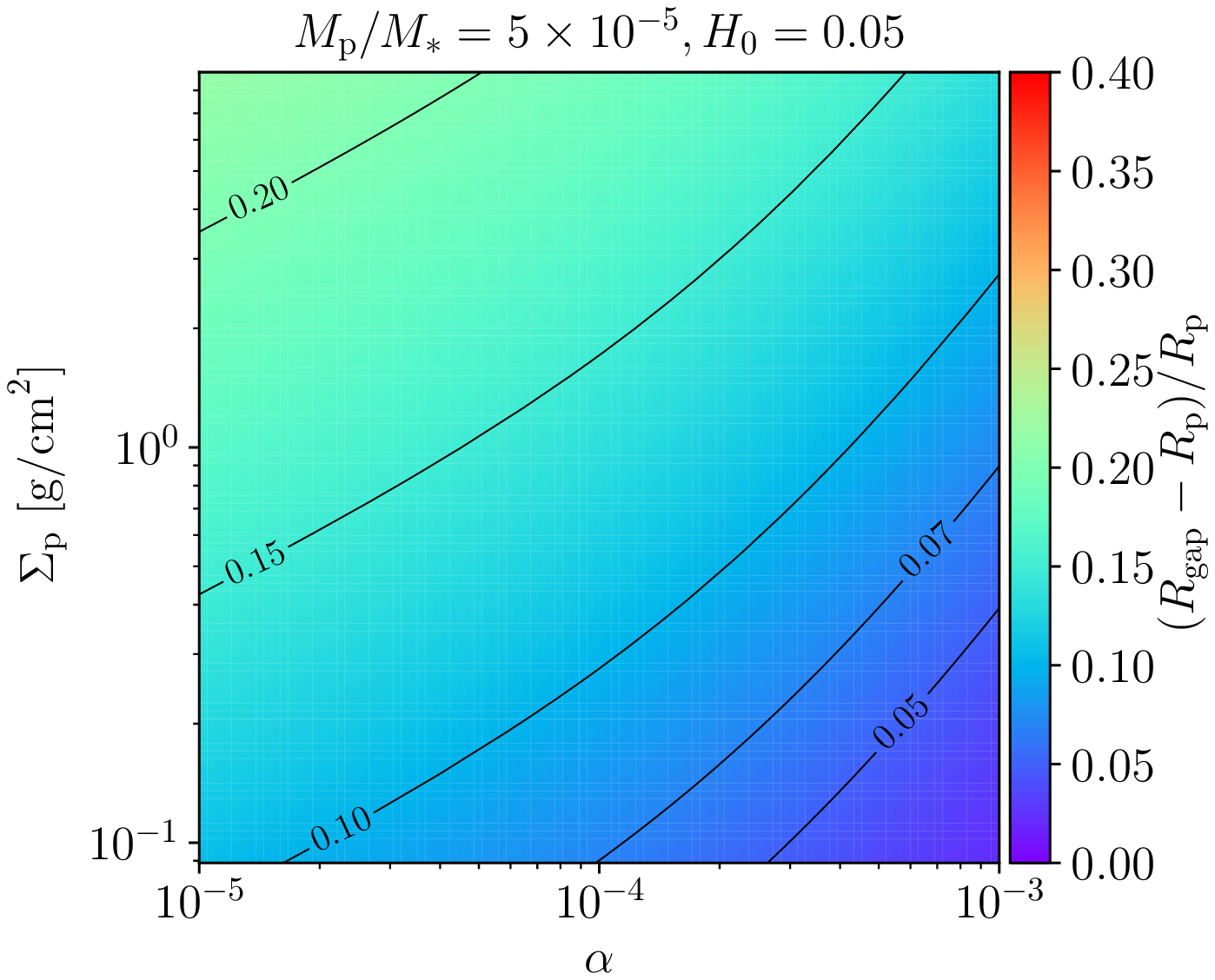}{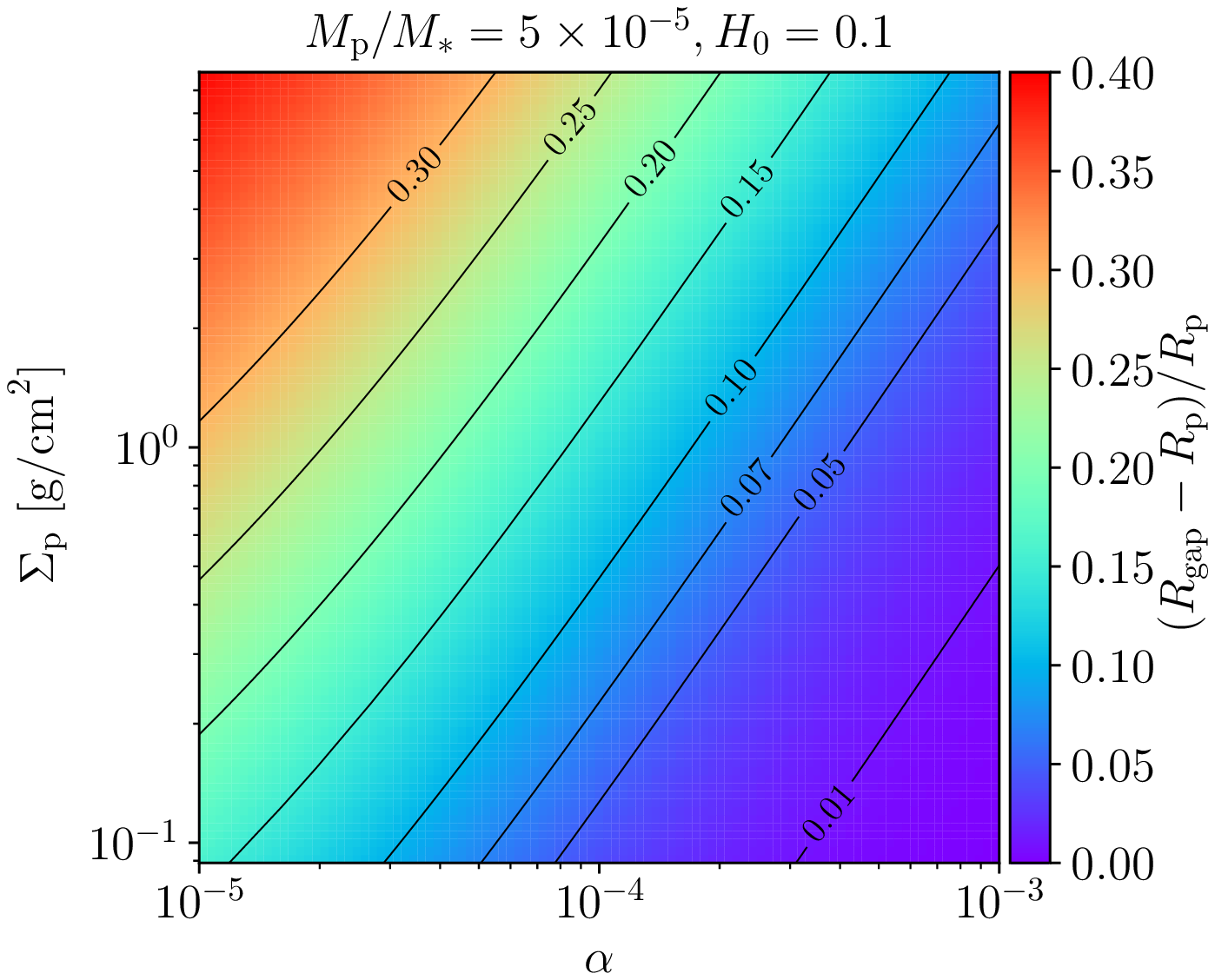}
\plottwo{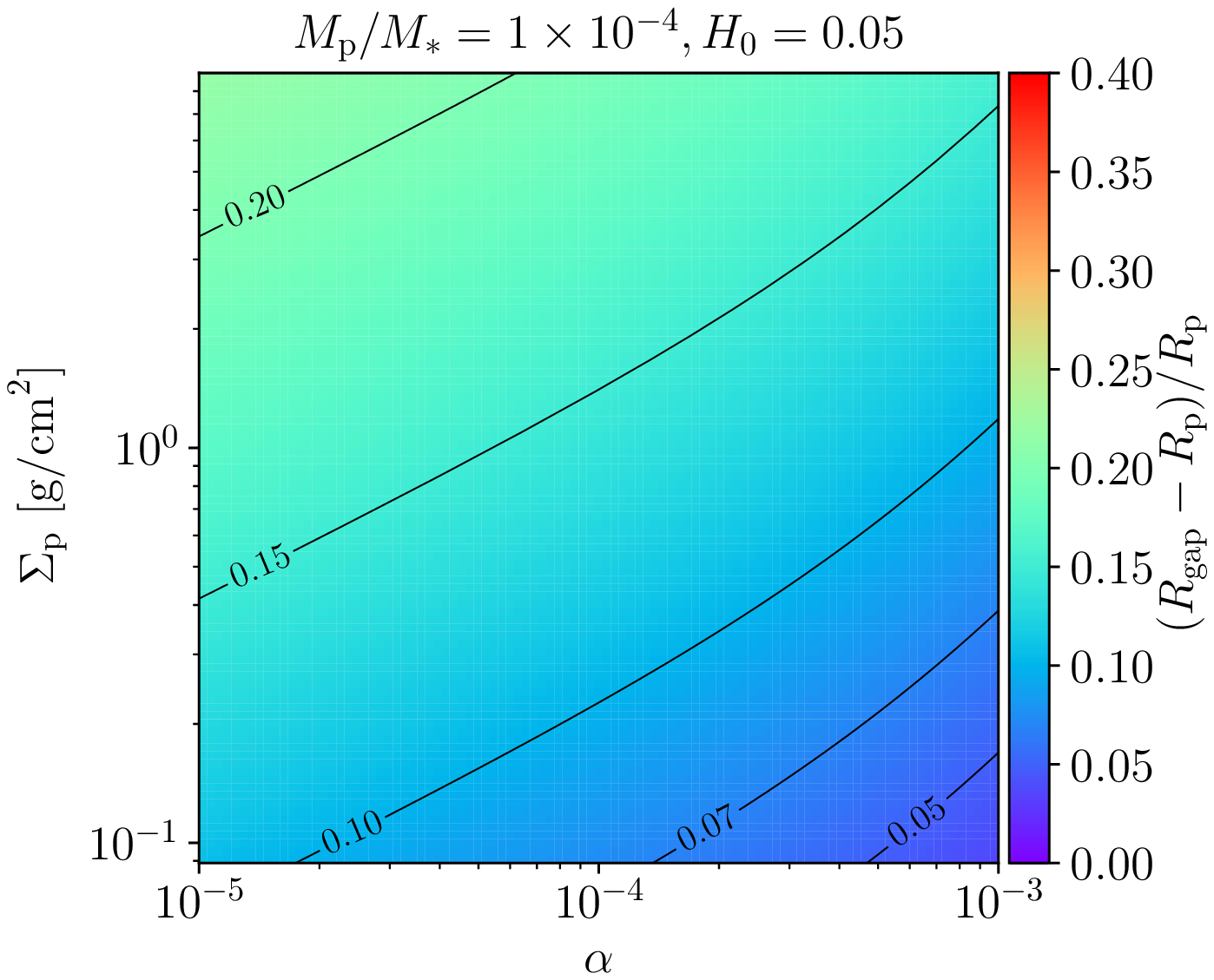}{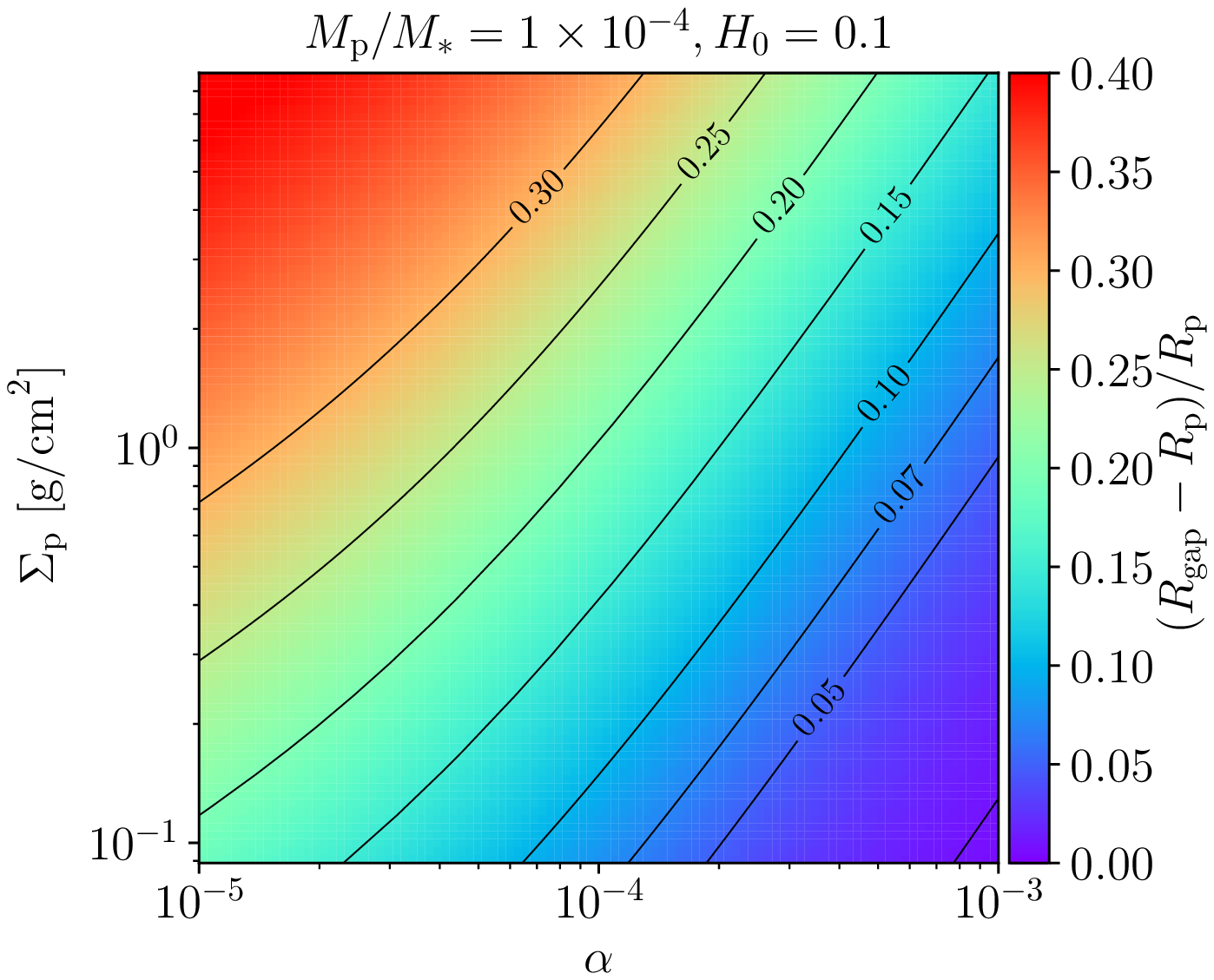}
\caption{
Radial difference between $\rp$ and $\rgap$ as a function of the viscosity $\alpha$ and the gas surface density $\sigmaunp$.
The planet mass is $\mpl/\mstar=5\times 10^{-5}$ (Neptune size) in the upper raw, and $\mpl/\mstar=1\times 10^{-4}$ in the lower raw, respectively, and we also assume $\mstar=1M_{\odot}$ and $R_0=100$~AU in the figures.
The disk aspect ratio is $0.05$ in the left column and $H_0=0.1$ in the right column, respectively.
\label{fig:obs_implication}}
\end{figure*}
Figure~\ref{fig:obs_implication} shows that the radial difference between $\rp$ and $\rgap$ as a function of $\alpha$ and $\sigmaunp$ with the given planet masses ($\mpl/\mstar=5\times 10^{-5}$ in the upper panels and $\mpl/\mstar=1\times 10^{-4}$ in the lower panels) and aspect ratios ($H_0=0.05$ in the left panels and $H_0=0.1$ in the right panels).
When the difference between $\rp$ and $\rgap$ is measured from the observation, we can constrain $\alpha$ and $\Sigma_0$ along the line corresponding to be the observed value of $\rp-\rgap$ in Figure~\ref{fig:obs_implication}.
As can be seen from the figure, the difference between $\rp$ and $\rgap$ is relatively sensitive on $H_0$, whereas it does not significantly depend on the mass of the planet.

In addition to the difference between $\rp$ and $\rgap$, we may constrain the viscosity and the disk aspect ratio from the secondary gap formed in the inner disk, as shown in Section~\ref{subsec:2ndgap}.
When the visible secondary gap is observed, we can give the upper limit of the $\alpha$-parameter, namely $\alpha \lesssim 3\times 10^{-4}$, in the vicinity of the planet.
When no secondary gap is observed, on the other hand, we can give the lower limit of the $\alpha$-parameter as $\alpha \gtrsim 3\times 10^{-4}$.
Moreover, if the secondary gap is observed, the aspect ratio can be estimated from the location of the secondary gap by Equation~(\ref{eq:location_2ndgap}), which is an independent constraint from that by dust/gas emissions.
With the upper/lower limit of $\alpha$ and the constraint of the disk aspect ratio from the depth and the location of the secondary gap, we can more accurately estimate the viscosity and, especially the gas surface density from Figure~\ref{fig:obs_implication}.

\subsection{Caveat of our model} \label{subsec:caveat}
In this paper, we adopt the simple locally isothermal equation of state (EoS).
However, recently \cite{Miranda_rafikov2019} shows that simulations with the locally isothermal EoS can overestimate the contrast of ring and gaps features, as compared with results given by simulations with adiabatic EoS, even when the adiabatic index is $1.001$.
As can be seen from Figure~2 of \cite{Miranda_rafikov2019}, this discrepancy becomes significant for the gap and ring structures formed by a relatively large dust grains ($St\gtrsim 0.01$).
For the gas structures, the location of the primary and secondary gaps do not change much between locally isothermal EoS and adiabatic EoS cases.
In this paper, we consider the primary and secondary gaps.
Hence, our results would not be significantly affected in the adiabatic disk with the adiabatic index being $1.001$.

In the adiabatic disk, the torque exerted on the planet (especially the horseshoe torque) can be different from that in the locally isothermal disk \citep[e.g.,][]{Paardekooper_Baruteau_Crida_Kley2010}.
The migration velocity of the planet in the adiabatic disk can be slower than that in the locally isothermal disk \citep[e.g.,][]{Bitsch_Johansen_Lambrechts_Morbidelli2015}. 
The non-isothermal effects may affect the gap structure, though it may not be significant \citep{Kley_Crida2008}.
However, in an outer region where is optically thin, the cooling can be efficient.
In this case, the isothermal EoS could be good approximation.

The torque exerted from the large dust grains (so-called pebbles) can significantly slow the inward planetary migration down due to an asymmetric distribution of pebbles, as discussed by \cite{Benitez_Pessah2019}. 
However, when the planet forms gap and the mass of the planet is larger than the so-call pebble-isolation mass \citep[e.g.,][]{Morbidelli_Nesvorny2012,Lambrechts_Johansen_Morbidelli2014,Bitsch_Morbidelli_Johansen_Lega_Lambrechts_Crida2018}, such large dust grains cannot reach the vicinity of the planet.
In this case, the planet hardly feels the torque exerted from the pebble.
When the mass of the planet is larger than the pebble-isolation mass, the pebbles accumulate at an outer edge of the gap.
Since the surface density of the gas at the outer edge decreases due to the feedback from the pebbles accumulated at the outer edge, the inward migration of the planet also significantly slows down or change a direction of the migration \citep{Kanagawa2019}.
However, this effects is significant when an amount of the pebbles are accumulated at the outer edge by catching up with the planet.
When the inward migration of the planet is fast, this effect may be inefficient since the relative speed between the pebble and the planet is not large enough.

If an actual migration velocity is deviated from that given by Equation~(\ref{eq:tmig_incomp}), we could overestimate/underestimate the surface density of gas around the planet.
This overestimate/underestimate could be found by comparing with the CO density estimated by the CO emission, though the CO density also has uncertainties related to e.g., CO/H ratio.

In the parameter range that we investigated in this paper, the planet migrates only inward.
However, several mechanisms discussed above may change the migration speed and let the planetary migration outward.
Even when the outward migration, the location of the planet and the gap could be shifted.
In this case, the planet would be detected at the outer edge of the gap and thus $\rp>\rgap$.

\section{Summary} \label{sec:summary}
We investigated effects of the fast inward migrating planet on the shape of the gap in the protoplanetary disk when both the planet and the gap are observed, by carrying out hydrodynamic simulations.
Our results are summarized as follows:
\begin{enumerate}
  \item we found that the orbital radius of the planet ($\rp$) can be shifted inward from the location of the gap ($\rgap$).
  When the radial shift between the locations of the planet and the gap is observed, it can be evidence that the planet is formed in the outer region and migrating to the inner region quickly.
  \item We also derived the empirical formula between the radial shift of $\rp$ and $\rgap$ and the ratio of the migration and gap-opening timescales (Equation~\ref{eq:empirical_fit}). 
  The radial difference between $\rp$ and $\rgap$ becomes larger as the migration timescale is shorter than the timescale of the gap-opening.
  \item Since the ratio of the timescales of the migration and the gap-opening is a function of the planet mass and disk parameters (gas surface density, aspect ratio, viscosity), we can constrain these quantities (especially the viscosity and the gas surface density) from the observation, by using Equation~(\ref{eq:empirical_fit}).
  \item When the viscosity is relatively low, the secondary gap can be formed in the inner disk. The depth and location of the secondary gap depends on the viscosity and the aspect ratio, respectively (Figures~\ref{fig:2ndGap_depth} and \ref{fig:2ndGap_location}). If the secondary gap is observed, we can constraint the viscosity as $\alpha \lesssim 3\times 10^{-4}$. Otherwise, we can obtain the lower limit of the viscosity as $\alpha \gtrsim 3\times 10^{-4}$. The secondary gap is formed in a more inner part with a larger disk aspect ratio (Equation~\ref{eq:location_2ndgap}). By using these constraints from the secondary gap, we can further estimate the parameters in the planet formation region.
\end{enumerate}

\acknowledgements
We would like to thank the anonymous referee for his/her constructive comments which are helpful to improve this paper.
KDK was also supported by JSPS Core-to-Core Program ``International Network of Planetary Sciences'' and JSPS KAKENHI grant 19K14779.
TM was supported by JSPS KAKENHI grant 17H01103, 18H05441, 19K03932.
Numerical computations were carried out on the Cray XC50 at the Center for Computational Astrophysics, National Astronomical Observatory of Japan and the computational cluster of Research Center for the Early Universe.
\software{FARGO \citep{Masset2000}, Matplotlib \citep[\url{http://matplotlib.org}]{Matplotlib}, NumPy \citep[\url{http://www.numpy.org}]{NumPy}}

\appendix
\section{Resolution dependence} \label{sec:reso_dep}
\begin{figure}
%\epsscale{0.6}
\plotone{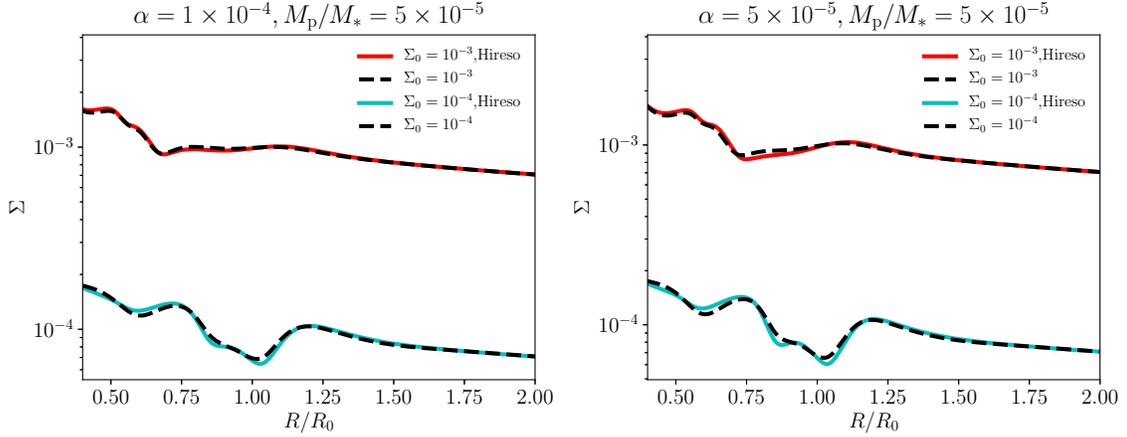}
\caption{
Azimuthally averaged surface density distributions at $t=1000\ t_0$, when $\mpl/\mstar=5\times 10^{-5}$, $H_0=0.05$.
In the left panel, $\alpha=1\times 10^{-4}$ and in the right panel, $\alpha=5\times 10^{-5}$.
The solid lines indicate the results given by the simulations with the high-resolution ($N_r=1024,N_{\phi}=2048$) and the dashed lines indicate the results given by the simulations with the standard resolution ($N_r=512,N_{\phi}=1024$).
\label{fig:comp_avgdens_with_HR}}
\end{figure}
In this appendix, we discuss resolution convergence of our results.
We carried out hydrodynamic simulations with higher resolution (1024 and 2048 meshes in radial and azimuthal directions, respectively) as compared with our standard resolution (512 and 1024 meshes in radial and azimuthal direction, respectively).
In Figure~\ref{fig:comp_avgdens_with_HR}, we compare the azimuthally averaged surface density at $t=1000\ t_0$ with the cases of the high-resolution and the standard resolution, when $\mpl/\mstar=5\times 10^{-5}$, $H_0=0.05$ and $\alpha=1\times 10^{-4}$ (left panel) and $\alpha=5\times 10^{-5}$ (right panel).
One can confirm that the surface density distributions are almost converged.

\begin{figure}
%\epsscale{0.6}
\plotone{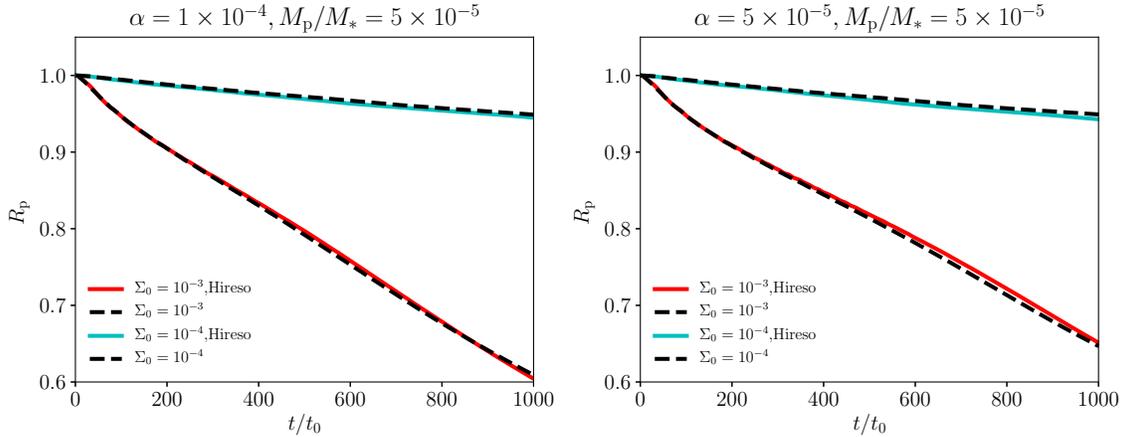}
\caption{
Comparison of evolution of the orbital radius of the planet given by the simulations with the high-resolution and standard resolution. 
The solid and dashed lines represent the results given by the simulations with the high-resolution and the standard resolution, respectively.
\label{fig:comp_orbitevo_with_HR}}
\end{figure}
In Figure~\ref{fig:comp_orbitevo_with_HR}, we compare the evolution of the orbital radius of the planet given by the simulations with the high-resolution and the standard resolution.
The evolution of the orbital radius is also quite similar to each other, in the cases of the high-resolution and the standard resolution.

\section{Location of the secondary gap} \label{sec:location_2ndgap}
\begin{figure}
%\epsscale{0.6}
\plotone{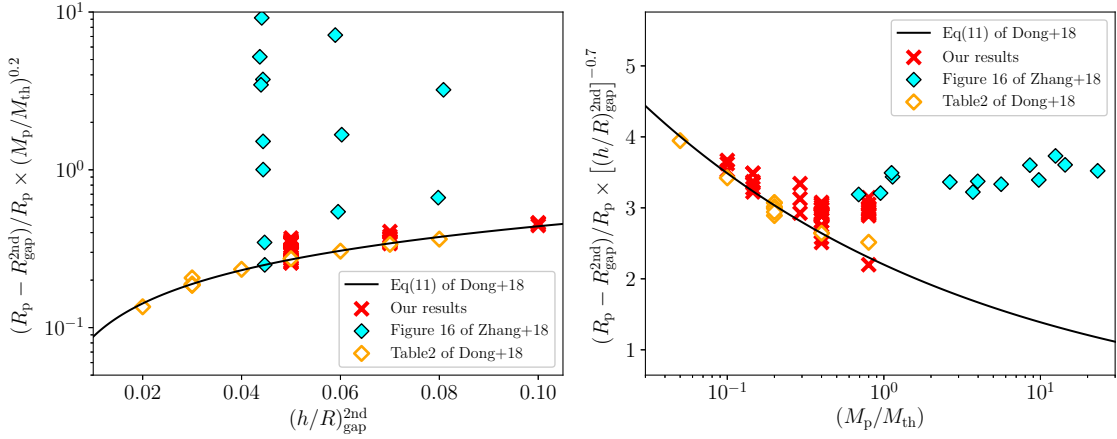}
\caption{
The same as Figure~\ref{fig:2ndGap_location_vs_aspectr}, but the axes are different.
Moreover, we plot all the data of \cite{Dong_Li_Chiang_Li2018} including these for $\mpl/M_{\rm th}<0.2$.
The red crosses, cyan diamonds, and orange diamonds represent the data given by this paper, \cite{Zhang_DSHARP2018} and \cite{Dong_Li_Chiang_Li2018}, respectively.
The solid line indicates the empirical formula given by \cite{Dong_Li_Chiang_Li2018}.
\label{fig:2ndgap_location_vs_aspectr2}}
\end{figure}
\cite{Dong_Li_Chiang_Li2018} obtains the empirical formula as (Equation~11 of that paper)
\begin{align}
\frac{\rgapsecond -\rp}{\rp} &= 0.27 \bracketfunc{\mpl}{M_{\rm th}}{-0.2} \bracketfuncb{(h/R)^{\rm 2nd}_{\rm gap}}{0.05}{0.7}.
\label{eq:formula_dong18}
\end{align}
Since $M_{\rm th} = (\hp/\rp)^3$, Equation~(\ref{eq:formula_dong18}) depends on $(h/R)^{1.3}$ (where we neglect spatial distribution of $h/R$ for simplicity).
In Figure~\ref{fig:2ndgap_location_vs_aspectr2}, we compare the data given by our simulations, \cite{Dong_Li_Chiang_Li2018} and \cite{Zhang_DSHARP2018} with Equation~(\ref{eq:formula_dong18}).
In the left panel, we show the dependence of $h/R$.
Equation~(\ref{eq:formula_dong18}) well reproduces the results of \cite{Dong_Li_Chiang_Li2018} and ours, but it does not match to the results given by \cite{Zhang_DSHARP2018}.
In the right panel of Figure~\ref{fig:2ndgap_location_vs_aspectr2}, we show the dependence of $\mpl/M_{\rm th}$.
Our results are consistent with the prediction given by Equation~(\ref{eq:formula_dong18}) when $\mpl/M_{\rm th} \lesssim 0.3$.
As $\mpl/M_{\rm th}$ increases, the results given by our simulations deviates from the prediction given by Equation~(\ref{eq:formula_dong18}).
Since \cite{Zhang_DSHARP2018} investigated a large planet, namely $\mpl/M_{\rm th} \gtrsim 1$, Equation~(\ref{eq:formula_dong18}) cannot reproduce these data.

%\bibliography{reference}
%\bibliographystyle{aasjournal}

\end{document}